# Establishing A Minimum Generic Skill Set
# For Risk Management Teaching
# In A Spreadsheet Training Course

Chadwick D.,
School of Computing and Mathematical Sciences,
University of Greenwich, London, SE10 9LS,UK
D.R.Chadwick@gre.ac.uk

## ABSTRACT

Past research shows that spreadsheet models are prone to such a high frequency of errors and data security implications that the risk management of spreadsheet development and spreadsheet use is of great importance to both industry and academia. The underlying rationale for this paper is that spreadsheet training courses should specifically address risk management in the development process both from a generic and a domain-specific viewpoint. This research specifically focuses on one of these namely those generic issues of risk management that should be present in a training course that attempts to meet good-practice within industry. A pilot questionnaire was constructed showing a possible minimum set of risk management issues and sent to academics and industry practitioners for feedback. The findings from this pilot survey will be used to refine the questionnaire for sending to a larger body of possible respondents. It is expected these findings will form the basis of a risk management teaching approach to be trialled in a number of selected ongoing spreadsheet training courses.

## 1. QUESTIONNAIRE AS A DATA GATHERING TECHNIQUE

The questionnaire reported upon herein was a pilot survey sent to ten persons of whom six responded. It was intended that the results from this would be used to create a fuller more comprehensive questionnaire for eventual sending to at least 100 possible respondents. To establish the skills that needed to be covered in generic spreadsheet risk management training the pilot questionnaire was constructed around a minimum set of 16 questions addressing five training methods to be used and eleven generic skills to be taught. The questionnaire was also constructed with facility for respondents to suggest further areas of concern that needed to be covered. For accuracy, the pilot survey had to simulate the eventual final questionnaire as closely as possible so its preamble, questions, and distribution method were carefully considered. For the theoretical research perspective and research paradigm see appendix C.

## 2. THE PILOT QUESTIONNAIRE

### 2.1 The Six Sections

1. **Pre-amble**: to explain the rationale of the questionnaire to the respondent.

2. **Generic Training Methods**: a sample of 5 generic training methods suggested by the author. Respondents were asked to comment upon these and rank their suitability for use in a training course by marking a 1-5 Likert scale.

3. **Generic Training Content**: a sample of 11 generic spreadsheet training content suggestions suggested by the author. Respondents were asked to comment upon these and rank their suitability for inclusion in a minimum generic skill set for a training course by marking a 1-5 Likert scale.

4. **'Anything else?' section** : five open sections were deliberately added to encourage respondents to add suggestions of their own. Such feedback considered essential as it was expected that not all pertinent issues had been addressed.

5. **'Some Information About You' section**: collecting data about the respondents themselves e.g. for respondents who were trainers:

| |
|---|
| Please identify what kind of a trainer you are : |
| What specific areas of modelling do you teach? |
| May I approach you again to discuss your answers? |

And for respondents from industry:

| |
|---|
| What industry are you involved with? |
| Do you think that spreadsheet training approaches should be improved? |

6. The questionnaire was ended with a completely open section for further comments.

**The initial 16 questions of the survey, the rationale for their inclusion, their drawbacks and supporting references are to be found in the table in Appendix A. The actual questionnaire is to be found in Appendix B.**

## 2.2 Presentation Of Questions

It was initially considered to present the questionnaire without an explanatory preamble as the reading of this would take up respondent time. However, a short preamble was eventually included to clearly set the scene for the potential respondent.

The two initial sections of the questionnaire were: 'Generic Training Methods' and 'Generic Training Content'. Each of these was included with an explanation of what these terms actually meant so misunderstandings could be limited – see example 1 below.

| GENERIC TRAINING METHODS | Aim: every course should have at least one instance of the following practices to raise student's awareness of error situations and to develop self-reflective practices. |
|---|---|

Example 1 : Explanation of section 'Generic Training Methods'

Similarly each question about a content or method to be considered was accompanied by a full descriptor – see example 2 below.

| Peer-audit (non-participative) | Student to find errors (if any) in another student's model |
|---|---|

Example 2 : Example of question descriptor for question 2

Respondents were encouraged to give their own suggestions in the 'Anything else?' sections of the questionnaire and in the last section entitled 'Please make any further comments below' – see Appendix B for the Actual Questionnaire sent out.

## 2.3 Scoring The Questionnaire

**2.3.1 Yes/No answers**:  were considered but discounted because **a** Yes/No answer, although easier to use for statistical purposes,  is too coarse a discriminator of opinion. Similarly, the use of all open questions, although possibly fruitful in new data, were also decided against as respondents may have found them too time consuming with resultant poor return response and/or a poor image of the survey.

**2.3.2 Likert scale** : was therefore adopted for scoring the questionnaire. Each question reply was given a weighting of importance by use of a Likert scale 1 - 5. Five grades were considered sufficient to obtain worthwhile discrimination – less would have been too coarse. A guide to answering was also shown e.g.

| 1=Not needed | 2 | 3=Indifferent | 4 | 5= Must have |
|---|---|---|---|---|

The end and middle points had concise explanations included : 'Not Needed' and 'Must Have' were diametric opposites and the mid-point was deliberately chosen as 'Indifferent' rather than left to open interpretation as say 'Ok but would leave out if something better came along'  or 'Ok but may be optional'.

**2.3.3 Free text** : a section was included at the end of the questionnaire were included to enable more open answers, suggestions and comments. Interpretation of open section responses was done carefully to avoid personal bias e.g. bias could be introduced by the author taking on board only those comments he liked and so giving a personal weighting to the importance of the comment.

## 3. ANALYSIS OF THE COMPLETED QUESTIONNAIRES

### 3.1 Ranking

The Likert scale made analysis fairly straightforward especially when results of the six respondents were placed into an Excel spreadsheet for analysis – see Appendix D.
Table 1 shows the final ranking of the sixteen issues in order of the totals of the Likert scale grading. This indicates the order of importance to the pilot survey's six respondents. It is clear from the responses that the issues mentioned in the questionnaire have different importance to the respondents  –  the use of Integral Documentation appears to be the most significant issue to be addressed with the teaching of a Taxonomy of Errors as the least important and by a wide margin.

### 3.2 Free Text Comments

In addition to the feedback on the initial 16 suggestions, the respondents also gave free text comments under the 'Anything else?' prompts on the questionnaire sheet. Some of these were useful and covered material initially overlooked – see Appendix E. These suggestions will be included as specific areas in the final more comprehensive questionnaire to be given to a wider audience at a later date.

### 3.3 Differences In Respondent Cohort

Interestingly, there was a slight but noticeable difference between the answers of the trainers cohort (Total Likert Score 223) and that of the business persons cohort (Total Likert Score 206) – see Appendix D. The latter appeared more cautious in giving a 5-rating (the highest) to any suggestion. It is not clear what this indicates – further research may be necessary. In addition, question 6 about inclusion of 'Taxonomy of common errors' had the most marked differential with trainers giving a total score of 13 against a score of 8 from the business persons.

| Original Question Number | Question Title | Question Descriptor | Total Likert Score |
|---|---|---|---|
| 13 | Integral Documentation approach | Documentation within the spreadsheet itself. | 30 |
| 1 | Error-seeded models | Student to find errors in tutor constructed model | 29 |
| 4 | Case Study | Student to discuss real-world models, and possible error situations | 29 |
| 10 | Auditing Tools (Integral) | Built-in audit functions i.e those integral to Excel | 29 |
| 12 | Access Control procedure | Password mechanisms etc | 29 |
| 2 | Peer-audit (non-participative) | Student to find errors (if any) in another student's model | 28 |
| 15 | Formulae Hard-coding controls | Guides to when hard-coding may be permitted and not permitted e.g universal constants in physics? | 28 |
| 7 | Spreadsheet engineering methodology | An stepped approach of some kind to aid a student in during spreadsheet building | 27 |
| 8 | Version Control approach | A structured approach to naming and storing past models | 27 |
| 16 | Named Ranges | When to be used or not | 27 |
| 5 | Self-Audit | Student checks own work, makes a statement as to how correct they think it is prior to tutor assessment | 26 |
| 14 | Formulae length limitations . | Heuristics to limit formulae length e.g. no formulae should have more than 8 operators | 26 |
| 9 | Confidentiality Controls | Spreadsheet encryption methods | 25 |
| 3 | Peer-audit (participative) | Student to find deliberately placed errors in another student's model | 24 |
| 11 | Auditing Tools (External) | Commercial audit tools and Excel add-ins e.g. OAK, SpACE | 24 |
| 6 | Taxonomy of common errors. | Classification of common errors that they can add their own errors to over time | 21 |

Table 1 : Questions sorted in order of Total Score on the Likert scale (from Appendix D).

## 4. CONCLUSION

The pilot questionnaire has given an indication of some of the generic skills of spreadsheet risk management that need to be included in a good-practice training course along with some of the training methods that should also be included. Not all the pertinent issues were mentioned in the original questionnaire as the free-text responses show. The results of the pilot questionnaire, along with ideas and suggestions from the free-text comments will be used to construct a more comprehensive questionnaire which will be sent to a larger set of potential respondents. These findings will in turn be used to establish a set of criteria for defining 'good-practice' in the training of spreadsheet risk management wherever this may occur.

# 5. REFERENCES

[1]     Butler R (2000) 'The Subversive Spreadsheet' Preface to Proceedings of EuSpRIG 2000 Conference, University of Greenwich, London, UK, July 2000

[2]     Chadwick D., et al (1999) "An Approach To The Teaching of Spreadsheets Using The Software Engineering Concept Of Modularisation" Proceedings of  INSPIRE '99, 4th Intern'l Conference on Software Process Improvement, Research, Education and Training, Heraklion, Crete 1999

[3]     Chadwick D.et al, 2000 'Quality Control in Spreadsheets: A Visual Approach using Colour Coding To Reduce Errors In Formulae' ;  Approaches to Software Quality Management , British Computer Society Conference, University of Greenwich, March 2000

[4]     Chadwick D., Sue R., 2001; 'Teaching  Spreadsheet Development Using Peer Audit And Self-Audit Methods For Reducing  Errors'   Proceedings of EuSpRIG 2001 Conference, Vrije Universiteit, Amsterdam, Holland, July 2001

[5]     Chadwick D., 2006; 'Education and Training Initiative : Supporting Learners and Teachers' Invited speaker at EuSpRIG 2006 Conference, University of Cambridge, Cambridge, UK, July 2006

[6]     Panko R. 2000; 'Spreadsheet Errors: What We Know, What We Think We Can Do' Proceedings of EuSpRIG 2000 Conference, University of Greenwich, London, UK , July 2000

[7]     Rogers A,; Teaching Adults; Open University Press, 2002

[8]     Chadwick D., et al  (1999) : The Ethical Problems Of Teaching Information Systems Security At Undergraduate Level; Proceedings of Ethicomp 99 Conference, Rome, Italy, July 1999

[9]     Blayney P.; 2006 ; 'Incidence and Effect of Errors Caused by Hard Coding of Input Data Values'; Proceedings of EuSpRIG 2006 Conference, University of Cambridge, Cambridge, UK, July 2006

[10]    EuSpRIG 2006 ; Yahoo Group Discussion Forum, Nov 2006; Formulae Length thread

[11]    Brunskell-Evans H.;2006, Ed.D programme teaching material;ACAD1061 Research: Theory and Methods 1, Avery Hill, University of Greenwich, Nov 2006.

[12]    Goddard W.;2006, Ed.D programme teaching material; ACAD1068 Curriculum Development, Avery Hill, University of Greenwich, Oct 2006.

# APPENDIX A: INITIAL QUESTION SUGGESTIONS : RATIONALE, DRAWBACKS AND REFERENCES

|  | SUGGESTIONS | RATIONALE | DRAWBACKS | REFER-ENCES |
|---|---|---|---|---|
| 1 | Error-seeded models e.g. student to find the errors | Game scenario. Challenging, competitive. | Student sees only as a game – does not draw inference to real life | [4], [5], [6] |
| 2 | Peer-audit (non-participative) e.g find the errors (if any) in another student's model | Both parties benefit. Modeller pays more attention, checks model prior to audit. | Modeller embarrassment if any error is found? | [4], [7] |
| 3 | Peer-audit (participative) e.g. find deliberately made errors in another student's model | Game, Challenging, Competitive. Both parties attentive to beat each other. | Auditor embarrassment if no error is found? | [4] |
| 4 | Case Study e.g of real-world models audited and found to be with errors. | Exposure to real-world model and possible mistakes. | Real-world models tend very complex and business domain specific. | [1], [6], [7] |
| 5 | Self-Audit | To grow self-awareness and reflection. | May become trivial so needs monitoring. | [4], [12] |
| 6 | Taxonomy of common errors. | Raise awareness of common errors. Student to add to it. | Problem of getting a meaningful taxonomy to start with. | [2], [5] |
| 7 | Spreadsheet engineering methodology | Give students a modelling process to follow. | Problem of getting a good methodology in the first place. | [2] |
| 8 | Version Control | Trail of who did what and when. | Hinders quick- and –dirty usage? | |
| 9 | Confidentiality Controls e.g. spreadsheet encryption | Confidentiality of data may be a legal necessity. | Encryption can be complex to explain. | [8] |
| 10 | Auditing Tools (Integral) | Learners need to be aware of audit functions. | Such audit functions are trivial and give false sense of security. | [1] |
| 11 | Auditing Tools (External) | Learners need exposure to commercial audit tools. | Whose products to choose? | [1] |
| 12 | Access Control e.g password mechanisms | Confidentiality of data may be a legal necessity | Hinders quick and easy use? | None |
| 13 | Integral Documentation | Gives useful metadata about the model. | Slow so either not done or not kept current. | None |
| 14 | Formulae length limitations | Long formulae are known to be a great source of error. | Splitting a formula may confuse the reader/user. | [10], [3] |
| 15 | Hard-coding of formulae controls e.g. guides to when hard-coding may be permitted | Hard coding of data known to be a great source of error when data needs to be changed. | Some constants need to be hard-coded? | [9], [3] |
| 16 | Named Ranges : use of | Known to have some advantages in clarifying structure. | ? | None |

# APPENDIX B: THE ACTUAL QUESTIONNAIRE : PAGE 1

The questionnaire pre-amble :

## Establishing A Minimum Generic Skill Set
## For Risk Management In A Spreadsheet Training Course.

Much research by Raymond Panko [2] and others has shown that human error is at the root of most spreadsheet errors. Other researchers [1] have suggested that human errors may be reduced by teaching not just 'how to do things correctly' but also 'how to avoid doing things incorrectly'.

I am researching what should be present in a 'good' training course that would help to reduce the known high frequency of human errors in spreadsheets. Such courses will be those provided by universities, private training organisations or company in-house. This questionnaire is a first attempt to identify those factors for the minimum set of training aims, methods and content that should appear in a good training course. Obviously there are generic and domain-dependent attributes for any training course. For instance we can, at this early stage, only identify those attributes that ALL courses should have regardless of the business area they may specifically be involved with.

This is a pilot – a first attempt at a questionnaire that will ultimately be sent to a wider audience. I am sending this to you as a possibly interested party and hope that you will participate by giving your views on what questions to ask, what factors to consider, and the weightings that should be given to different criteria.

If you would be interested in being more involved in this research please make contact. If we can establish some agreed approaches it may be possible for EuSpRIG to establish some good practice standards on training and perhaps, in time, accredit the courses of training providers meeting the agreed criteria.

Many thanks for your help with this.

[1]. Chadwick D. *Information Integrity In End-user Systems*. Chapman & Hall (Proceedings of the First Annual IFIP TC-11 Working Group 11.5 Working Conference on Integrity and Internal Control in Information Systems, Zurich, Switzerland 3-4 December 1997)

[2] Panko R. 2000; 'Spreadsheet Errors: What We Know, What We Think We Can Do' Proceedings of EuSpRIG 2000 Conference, University of Greenwich, London, UK

Please return to David Chadwick on cd02@gre.ac.uk

## Questionnaire follows on the next two pages.

# APPENDIX B: THE ACTUAL QUESTIONNAIRE: PAGE 2

**PART 1 of 2 : Establishing Criteria For A Good Training Course**

Please answer the following (circle your chosen answer):

| | GENERIC TRAINING METHODS | Aim: every course should have at least one instance of the following practices to raise student's awareness of error situations and to develop self-reflective practices. | 1=Not needed 2 3=Indifferent 4 5= Must have | | | | |
|---|---|---|---|---|---|---|---|
| 1 | Error-seeded models | Student to find errors in tutor constructed model | 1 | 2 | 3 | 4 | 5 |
| 2 | Peer-audit (non-participative) | Student to find errors (if any) in another student's model | 1 | 2 | 3 | 4 | 5 |
| 3 | Peer-audit (participative) | Student to find deliberately placed errors in another student's model | 1 | 2 | 3 | 4 | 5 |
| 4 | Case Study | Student to discuss real-world models, and possible error situations | 1 | 2 | 3 | 4 | 5 |
| 5 | Self-Audit | Student checks own work, makes a statement as to how correct they think it is prior to tutor assessment | 1 | 2 | 3 | 4 | 5 |
| | GENERIC TRAINING CONTENT | Aim: every student should be taught the following to give structure to the more domain-specific learning content | | | | | |
| 6 | Taxonomy of common errors. | Classification of common errors that they can add their own errors to over time | 1 | 2 | 3 | 4 | 5 |
| 7 | Spreadsheet engineering methodology | An stepped approach of some kind to aid a student in during spreadsheet building | 1 | 2 | 3 | 4 | 5 |
| 8 | Version Control approach | A structured approach to naming and storing past models | 1 | 2 | 3 | 4 | 5 |
| 9 | Confidentiality Controls | Spreadsheet encryption methods | 1 | 2 | 3 | 4 | 5 |
| 10 | Auditing Tools (Integral) | Built-in audit functions i.e. those integral to Excel | 1 | 2 | 3 | 4 | 5 |
| 11 | Auditing Tools (External) | Commercial audit tools and Excel add-ins e.g. OAK, SpACE | 1 | 2 | 3 | 4 | 5 |
| 12 | Access Control procedure | Password mechanisms etc | 1 | 2 | 3 | 4 | 5 |
| 13 | Integral Documentation approach | Documentation within the spreadsheet itself. | 1 | 2 | 3 | 4 | 5 |
| 14 | Formulae length limitations . | Heuristics to limit formulae length e.g. no formulae should have more than 8 operators | 1 | 2 | 3 | 4 | 5 |
| 15 | Formulae Hard-coding controls | Guides to when hard-coding may be permitted and not permitted e.g. universal constants in physics? | 1 | 2 | 3 | 4 | 5 |
| 16 | Named Ranges | When to be used or not | 1 | 2 | 3 | 4 | 5 |
| 17 | Anything else? | | 1 | 2 | 3 | 4 | 5 |
| 18 | Anything else? | | 1 | 2 | 3 | 4 | 5 |
| 19 | Anything else? | | 1 | 2 | 3 | 4 | 5 |
| 20 | Anything else? | | 1 | 2 | 3 | 4 | 5 |

# APPENDIX B: THE ACTUAL QUESTIONNAIRE : PAGE 3

**PART 2 of 2 : Some Information About You**

If you are a Training Provider
Please answer the following (circle your chosen answer):

| | |
|---|---|
| Please identify what kind of a trainer you are: | University    Private Trainer    Company In-house    Other<br><br>If other please state: |
| What specific areas of modelling do you teach? | General    Accounting    Engineering    Medicine    Other<br><br>If other please state: |
| May I approach you again to discuss your answers? | Yes         No |
| | |

**I**f you are not a Training provider
Please answer the following (circle your chosen answer)::

| | |
|---|---|
| What industry are you involved with? | Finance         Accounting         Engineering<br><br>Medicine         General         Other |
| Do you think that spreadsheet training approaches should be improved? | Yes         No<br><br>Please give your reasons: … |

The questionnaire is finished – again, many thanks for your help.

Please make any further comments below:

# APPENDIX C: Theoretical Research Perspective

The theoretical research perspectives were a combination of both Positivist and Interpretative paradigms [11].

A Positivist approach was partly adopted for three reasons:

1. A deliberate policy was adopted of making the questionnaire straightforward to answer because the respondents (especially business managers) would have limited time.
2. A positivist approach produces data that is amenable to tabular presentation. This was important so that findings could be clearly presented to all the respondents.
3. The respondent parties would be business consultants, in-house business trainers, private trainers and academic educators. Most of these would be cognizant with the positivist approach (especially the business people whose support was dearly needed).

An element of Interpretative approach was also required for open sections of the questionnaire where respondent suggestions were required. This was problematic in that interpretation was open to subjectivity but was necessary as it was possible that not all pertinent factors may have been covered in the initial survey questions (see paragraph 2.3 'Scoring The Questionnaire').

# APPENDIX D: ANALYSIS OF QUESTIONNAIRE RESULTS

| | AT: Academic Trainer PT: Private Trainer | | | | BC: Business Commercial | | | |
|---|---|---|---|---|---|---|---|---|
| Survey Question Number | AT Reply1 | AT Reply2 | PT Reply3 | Total | BC Reply4 | BC Reply5 | BC Reply6 | Total |
| 1 | 5 | 5 | 5 | 15 | 5 | 5 | 4 | 14 |
| 2 | 4 | 5 | 4 | 13 | 5 | 5 | 5 | 15 |
| 3 | 3 | 5 | 3 | 11 | 3 | 5 | 5 | 13 |
| 4 | 5 | 5 | 5 | 15 | 5 | 5 | 4 | 14 |
| 5 | 5 | 3 | 5 | 13 | 5 | 5 | 3 | 13 |
| 6 | 5 | 5 | 3 | 13 | 3 | 2 | 3 | 8 |
| 7 | 5 | 5 | 5 | 15 | 5 | 4 | 3 | 12 |
| 8 | 5 | 5 | 5 | 15 | 5 | 4 | 3 | 12 |
| 9 | 5 | 5 | 4 | 14 | 4 | 3 | 4 | 11 |
| 10 | 5 | 5 | 5 | 15 | 5 | 4 | 5 | 14 |
| 11 | 4 | 5 | 3 | 12 | 4 | 3 | 5 | 12 |
| 12 | 5 | 5 | 5 | 15 | 5 | 4 | 5 | 14 |
| 13 | 5 | 5 | 5 | 15 | 5 | 5 | 5 | 15 |
| 14 | 5 | 5 | 3 | 13 | 5 | 5 | 3 | 13 |
| 15 | 5 | 5 | 5 | 15 | 5 | 5 | 3 | 13 |
| 16 | 5 | 4 | 5 | 14 | 5 | 4 | 4 | 13 |
| Total | 76 | 77 | 70 | 223 | 74 | 68 | 64 | 206 |
| Out of | 80 | 80 | 80 | 240 | 80 | 80 | 80 | 240 |

| Survey Question Number | Total Likert Score | | Survey Question Number | Total Likert Score | Rank Position |
|---|---|---|---|---|---|
| 1 | 29 | | 13 | 30 | 1 |
| 2 | 28 | | 1 | 29 | 2 |
| 3 | 24 | | 4 | 29 | 3 |
| 4 | 29 | | 10 | 29 | 4 |
| 5 | 26 | | 12 | 29 | 5 |
| 6 | 21 | | 2 | 28 | 6 |
| 7 | 27 | | 15 | 28 | 7 |
| 8 | 27 | | 7 | 27 | 8 |
| 9 | 25 | | 8 | 27 | 9 |
| 10 | 29 | | 16 | 27 | 10 |
| 11 | 24 | | 5 | 26 | 11 |
| 12 | 29 | | 14 | 26 | 12 |
| 13 | 30 | | 9 | 25 | 13 |
| 14 | 26 | | 3 | 24 | 14 |
| 15 | 28 | | 11 | 24 | 15 |
| 16 | 27 | | 6 | 21 | 16 |

# APPENDIX E : SAMPLE OF FREE TEXT COMMENTS

The issues raised in the questionnaire are all important. However, the crucial issue is, how much time one has in a course.

Students have to understand that
- neat layout does not guarantee correctness, but poor layout is a good hint for incorrectness;
- certain working conditions will trigger faulty results
- every thing produced needs to be checked (hence, they should learn about techniques for checking spreadsheets).

Context of spreadsheet based decision-making – Why are they important? What might impact of correct (or incorrect) spreadsheet models be? – Vital to tell them WHY this stuff is important, easy for us old lags to forget and assume we're preaching to the choir when we're not.

Just because a syllabus supplied by a training organization contains an item does not mean that the individual trainer on the day effectively teaches it, or that the student learns it. A topic can appear on a course to get accreditation, but never be examined on; or if it appears on the test, is a question that is always skipped. Is it possible to establish criteria that allow us to examine actual training results to verify that these standards have been learnt?